\documentclass[journal=ancac3,manuscript=article]{achemso}
\usepackage{amssymb}
\usepackage{empheq,etoolbox,amsmath}
\patchcmd{\subequations}
  {\theparentequation\alph{equation}}
  {\theparentequation.\alph{equation}}
  {}{}

\usepackage{chemformula} 
\usepackage[T1]{fontenc} 

\author{F. Paolucci}
\email{federico.paolucci@unipi.it}
\affiliation{These authors equally contributed to this work.}
\affiliation{Dipartimento di Fisica "E. Fermi", Università di Pisa, Largo Bruno Pontecorvo 3, I-56127, Pisa, Italy}
\affiliation{INFN Sezione di Pisa, Largo Bruno Pontecorvo 3, I-56127 Pisa, Italy}
\author{F. Bianco}
\affiliation{These authors equally contributed to this work.}
\affiliation{NEST, Istituto Nanoscienze-CNR, Piazza San Silvestro 22, I-56127 Pisa, Italy}
\author{F. Giazotto}
\affiliation{NEST, Istituto Nanoscienze-CNR and Scuola Normale Superiore, Piazza San Silvestro 22, I-56127 Pisa, Italy}
\author{S. Roddaro}
\email{stefano.roddaro@unipi.it}
\affiliation{Dipartimento di Fisica "E. Fermi", Università di Pisa, Largo Bruno Pontecorvo 3, I-56127, Pisa, Italy}

\title{Active electron cooling of graphene}


\begin{document}

\begin{tocentry}
\includegraphics {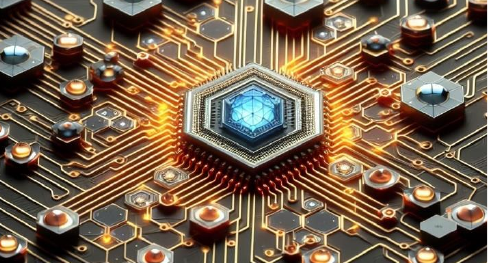}
\end{tocentry}


\begin{abstract}

In the emergent field of quantum technology, the ability to manage heat at the nanoscale and in cryogenic conditions is crucial for enhancing device performance in terms of noise, coherence, and sensitivity. Here, we demonstrate the active cooling and refrigeration of the electron gas in graphene, by taking advantage of nanoscale superconductive tunnel contacts able to pump or extract heat directly from the electrons in the device. Our structures achieved a top \emph{cooling} of electrons in graphene of $\sim15.5\,{\rm mK}$ at a bath temperature of $\sim 448\,{\rm mK}$, demonstrating the viability of the proposed device architecture. Our experimental findings are backed by a detailed thermal model that accurately replicated the observed behavior. Alternative cooling schemes and perspectives are discussed in light of the reported results. Finally, graphene electron cooling could find application in superconducting hybrid quantum technologies, such as radiation detectors, logic gates and qubits.

\end{abstract}


\section{Introduction}
Heat management in nano-scale solid state systems at cryogenic temperatures \cite{Fornieri2017-bn,Roddaro2011-zh,Fornieri2014-de,Guarcello2018-xk} is the cornerstone for quantum thermodynamics \cite{LIEB19991,Horodecki2013-cy}, and it strongly impacts superconducting and hybrid quantum technologies \cite{Gao2021,Acin2018-qy}.
Indeed, the performance of superconducting quantum devices can be improved by locally decreasing the electronic temperature. 
In this context, it is possible to distinguish between \emph{cooling}, i.e. leading the electrons colder than the crystal, and \emph{refrigeration}, i.e. actively decreasing the electronic temperature from a maximum to a minimum both larger than the lattice temperature. 
In particular, local electron cooling in hybrid superconducting systems opens the way to the creation of solid-state heat engines \cite{Germanese2022-ur}, by electrical control of the thermal cycles, and to an increase the qubit lifetime \cite{Verjauw2022-xd}, by mastering the quasiparticle population. In addition, direct electron cooling can be a great tool for improving the noise, coherence, and sensitivity of superconducting electronics \cite{Giazotto2005-fp,Tirelli2008-bd}, magnetometers, or radiation detectors. Indeed, voltage-biased superconducting tunnel junctions are routinely employed to decrease the electronic temperature in normal metals (NIS coolers) \cite{Nahum1994} and superconductors (SIS' coolers) \cite{Quaranta2011}.
This capability has been exploited to improve the sensitivity of transition-edge sensors (TESs) \cite{Miller2008} and to develop the cold-electron bolometers (CEBs) \cite{KUZMIN20002129}. For instance, CEBs showed experimental values of the noise equivalent power (NEP) as low as $3\times 10^{-18}$ W\,Hz$^{-1/2}$ \cite{Kuzmin2019-ru}, by also leveraging the damping of the electron-phonon thermalization at cryogenic temperatures \cite{Giazotto2006}.

In recent years, hybrid superconductor/semiconductor systems gained significant attention in the context of quantum technologies \cite{Benito2020}, since they marry the macroscopic phase-coherent nature of superconductors with the field-effect tunability of semiconductors. In particular, proximitized nanowires \cite{Casparis2016}, two-dimensional electron gases (2DEGs) \cite{Casparis2018-sj} and graphene \cite{Kroll2018-ue,Wang2019-uf} have been proposed as the fundamental building blocks of future \emph{gatemon} qubits.
Furthermore, superconducting elements coupled to semiconductor quantum dots \cite{Komiyama2000-eh}, nanowires \cite{Schapers1999} and graphene \cite{Lee2020-sl,Evan2021} have also shown to have promising perspectives as bolometers and single-photon detectors. Among all these hybrid superconductor/semiconductor systems, electron refrigeration was only demonstrated in (i) clean 2DEG/superconductor Josephson junction \cite{Gunnarsson2015-vf} and (ii) InAs nanowire/Al tunnel junctions \cite{Mastomaki2017-mv}.
For what concerns graphene, heat management has only been experimentally achieved in terms of electron-phonon thermalization controlled by overheating the quasiparticles with respect to the lattice, in the limit of ballistic, diffusive and suspended \cite{Betz2013-yh,Laitinen2014-cf} graphene. Despite theoretical studies indicating that graphene-based CEBs achieve very high efficiencies \cite{Vischi2020}, electron cooling has not been demonstrated due to the complexities of the fabrication of high-quality and low-resistance graphene/superconductor tunnel junctions \cite{Bretheau2017,Yu2011}.

\begin{figure*} [t!]
    \begin{center}
    \includegraphics {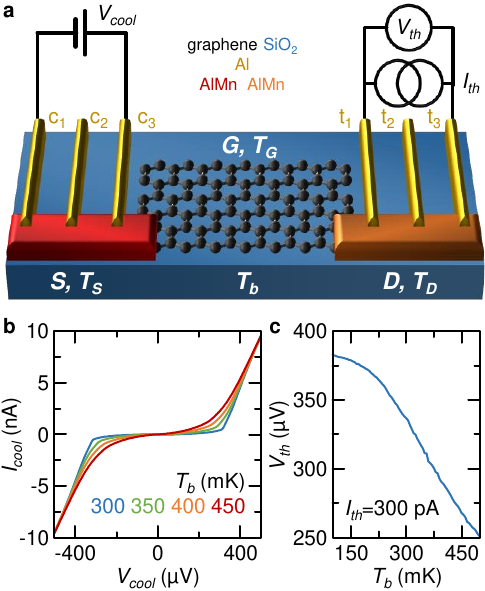}
    \end{center}
    \caption{\textbf{Schematics and basic characterization of the graphene electron cooler.}
    \textbf{a} Schematic representation of the device. The electronic temperature ($T_S$) of the metallic Al$_{0.98}$Mn$_{0.02}$ source electrode ($S$, red) is decreased/increased with respect to the substrate phonon temperature ($T_b$) by voltage biasing ($V_{cool}$) a couple of superconducting Al tunnel coolers ($c_1$ and $c_3$, yellow). The temperature ($T_D$) of the Al$_{0.98}$Mn$_{0.02}$ drain electrode ($D$, orange) is measured by current biasing ($I_{th}$) a couple of superconducting tunnel thermometers ($t_1$ and $t_3$, yellow) while recording the voltage drop ($V_{th}$). The measured modulation of $T_D$ is due to the change of the graphene electronic temperature ($T_G$).
    \textbf{b} Current ($I_{cool}$) versus voltage ($V_{cool}$) characteristics of the cooling tunnel junctions ($c_1$ and $c_3$) for selected values of the bath temperature ($T_b$).
    \textbf{c} Calibration of the thermometers ($t_1$ and $t_3$) in the form of the voltage ($V_{th}$) versus $T_b$ characteristics at constant current bias $I_{th}=300\,{\rm pA}$.}
\label{Fig1}
\end{figure*}

Here, we demonstrate graphene electron \emph{refrigeration} and net \emph{cooling} by exploiting superconducting NIS tunnel coolers. 
To this scope, we designed a thermal device, where a graphene flake is in good galvanic contact with metallic source and drain electrodes. 
The source electrode is equipped with tunnel coolers thus operating as a \emph{non-local} electron refrigerator for graphene thanks to the efficient heat diffusion across the whole device.
Indeed, the drain electrode is equipped with local thermometers to monitor the electron temperature throughout the device. 
This geometry allows us to take advantage of the high-quality tunnel barriers achievable in fully metallic junctions \cite{Giazotto2006}, thereby maximizing the device cooling efficiency. 
As a key technical progress towards the demonstration of active cooling of graphene electrons, we developed a complex fabrication protocol embracing the typical methods of two-dimensional materials and the shadow-mask evaporation technology of fully metallic superconducting devices \cite{Dolan}. In our experiments, we measured electron refrigeration of the drain electrode for a wide range of bath temperatures by voltage-biasing a couple of coolers at the source, demonstrating a best drain cooling of about $3\,{\rm mK}$ starting from $448\,{\rm mK}$. To extract the electronic temperature in the graphene sheet, we developed a theoretical model accounting for all the main thermal exchange mechanisms in the device and able to reproduce the experimental data with impressive accuracy. The resulting maximum cooling in graphene was $(15.5\pm0.5)\,{\rm mK}$ at a bath temperature of $448\,{\rm mK}$, with a best voltage-to-temperature transfer function of $(-115\pm4)\,{\rm K/ V}$ at $307\,{\rm mK}$. These results demonstrate that \emph{non-local} electron cooling is a promising strategy for reducing quasiparticle poisoning and noise in hybrid graphene quantum technologies, such as graphene-based coherent electronics, qubits, and radiation detectors \cite{Vischi2020}. 

\section{Results}
\subsection{Device design and basic characterization}
The device developed to demonstrate electron cooling in graphene is shown in Fig.~\ref{Fig1}a. 
The realization of this architecture was only possible by introducing the angle controlled shadow-mask evaporation technology \cite{Dolan} in the fabrication protocol of two-dimensional crystal-based devices realized on small chips. In fact, angle resolved evaporation requires homogeneous resist thin films that are usually spin-coated to entire wafers, as explained in the Experimental Section.
In our fabrication process, the mechanically exfoliated monolayer graphene flakes were shaped by means of electron-beam lithography (EBL) and reactive ion etching (RIE), while source ($S$, red), drain ($D$, orange), coolers ($c_i$ with $i=1,2,3$, yellow) and thermometers ($t_i$ with $i=1,2,3$, yellow) were realized by EBL and shadow mask evaporation through a suspended resist mask \cite{Dolan}. The details of the fabrication procedure are presented in the Experimental Section.

The rectangular ($8\times3\,{\rm \mu m^2}$) graphene flake ($G$, grey honeycomb), with electronic temperature $T_G$, is placed onto an intrinsic silicon substrate covered with a $300\,{\rm nm}$-thick silicon dioxide layer (SiO$_2$, blue) with phonon temperature $T_b$. To tune $T_G$, the graphene is connected through a clean galvanic contact to a normal metal electrode $S$ made of Al$_{0.98}$Mn$_{0.02}$ (the presence of Mn fully suppresses superconductivity), whose temperature ($T_S$) can be controlled by three tunnel-coupled superconducting aluminum fingers ($c_i$ with $i=1,2,3$, yellow). 
Indeed, any pair of superconducting tunnel leads can act as a SINIS electron refrigerator for a normal metal island when voltage biased near the sum of their superconducting energy gaps \cite{Nahum1994}. Differently, the same electrodes act as local heaters when biased in the normal state. Consequently, $T_G$ can be tuned by a voltage ($V_{cool}$) applied to a pair of cooling junctions thanks to the cold/hot electron diffusion from $S$ to $G$. 
To extract the basic properties of all the tunnel coolers, we investigated their transport properties in a two-terminal configuration, as shown in Fig.~\ref{Fig1}b for $c_1$ and $c_3$ at different values of $T_b$. In particular, we extracted their normal-state resistance ($R_{cool_1}\simeq3.6\,{\rm k\Omega}$, $R_{cool_2}\simeq22\,{\rm k\Omega}$ and $R_{cool_3}\simeq17.8\,{\rm k\Omega}$) and zero-temperature superconducting energy gap ($\Delta_{0,cool_1}\simeq175\,{\rm \mu eV}$, $\Delta_{0,cool_2}\simeq190\,{\rm \mu eV}$ and $\Delta_{0,cool_3}\simeq190\,{\rm \mu eV}$). We note that, according to the low value of $R_{cool_1}$, the energy gap of $c_1$ is lower than that of the other leads since it might be subject to a sizeable inverse proximity effect from $S$ \cite{DeGennes}. To ensure large cooling power and homogeneity of $T_S$ \cite{Nahum1994,Giazotto2006}, we chose $c_1$ and $c_3$ as electronic coolers for our graphene flake. Differently, $c_2$ will only act as a thermal loss to the phonon bath.

The changes of $T_G$ are non-locally detected by measuring the electronic temperature ($T_D$) of an Al$_{0.98}$Mn$_{0.02}$ metallic drain electrode ($D$, orange) placed in good electrical contact with $G$. Indeed, $T_D$ can be probed by current ($I_{th}$) biasing a couple of superconducting tunnel leads ($t_i$ with $i=1,2,3$, yellow) and recording the voltage drop ($V_{th}$) \cite{Rowell1976}. 
Indeed, since the quasiparticle transport of a SINIS structure depends mainly on the electronic temperature of the metallic island \cite{Giazotto2006}, $V_{th}$ is the perfect parameter to determine $T_D$.
Since the three electronic thermometers show very similar properties (see Experimental Section for details), we chose to exploit $t_1$ and $t_3$ to sample the temperature over the largest possible portion of $D$. Figure~\ref{Fig1}c shows the calibration curve of the thermometers ($V_{th}$) as a function of $T_b$ obtained by injecting $I_{th}=300\,{\rm pA}$.

\begin{figure} [t!]
    \begin{center}
    \includegraphics {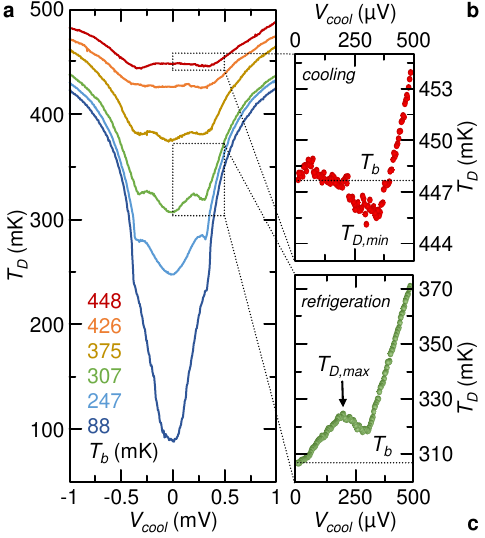}
    \end{center}
    \caption{\textbf{Temperature modulation of the drain electrode.}
    \textbf{a} Electronic temperature of the drain electrode ($T_D$) as a function of the voltage bias of the source coolers ($V_{cool}$) recorded at selected values of bath temperature ($T_b$).
    \textbf{b} $T_D$ versus $V_{cool}$ characteristics recorded at $T_b=448\,{\rm mK}$. The system shows \emph{cooling}, that is $T_D<T_b$ for specific values of $V_{cool}$. $T_{D,min}$ represents the minimum recorded value of $T_D$.
    \textbf{c} $T_D$ versus $V_{cool}$ characteristics recorded at $T_b=307\,{\rm mK}$. The system shows \emph{refrigeration}: the coolers are able to decrease the drain temperature from a maximum value ($T_{D,max}$), but  $T_D>T_b$ always applies.}
\label{Fig2}
\end{figure}

\subsection{Active cooling experiments}
The experimental set-up for the demonstration of active electron cooling in graphene is sketched in Fig.~\ref{Fig1}a. The measurements were performed by sweeping $V_{cool}$ while recording $T_D$. To probe both cooling (for 
$|V_{cool}\lesssim(\Delta_{cool_1}+\Delta_{cool_3})/e$, with $e$ the electron charge) and heating (for $|V_{cool}|>(\Delta_{cool_1}+\Delta_{cool_3})/e$), $V_{cool}$ was swept from $-1\,$mV to $1\,$mV. Furthermore, we determined the impact of phonon temperature by stabilizing the cryostat at desired values of $T_b$.
Figure~\ref{Fig2}a shows the modulation of $T_D$ with $V_{cool}$ recorded for selected values of $T_b$. At low values of bath temperature, $T_D$ monotonically increases with $|V_{cool}|$, as shown from the blue trace measured at $T_b=88\,{\rm mK}$, thus the source serves only as an electron heater. At higher values of $T_b$, the electronic temperature of $D$ is non-monotonic for a range of $V_{cool}$ that widens with increasing $T_b$.
The device shows two distinct operation conditions depending on the bath temperature: in the \emph{cooling} regime $T_D$ is lower than $T_b$ for certain values of $V_{cool}$ (see Fig.~\ref{Fig2}b), while in the \emph{refrigeration} regime $T_D$ decreases compared to a maximum value ($T_{D,max}$) by increasing $V_{cool}$ but it is always higher than $T_b$ (see Fig.~\ref{Fig2}c). In particular, the drain electrode shows a maximum cooling of $T_{b}-T_{D,min}\simeq3\,{\rm mK}$ at $T_b=448\,{\rm mK}$ and the best refrigeration of $T_{D,max}-T_{D,min}=(15.5\pm0.5)\,{\rm mK}$ (with $T_{D,min}$ the minimum temperature recorded) at $T_b=307\,{\rm mK}$.

The low-voltage electron overheating shown by every experimental curve in Fig.~\ref{Fig2}a is due to the significant number of under-gap quasiparticle states (described by the Dynes broadening parameter $\Gamma$ \cite{Dynes1984}, see Methods) present in the cooler tunnel junctions. In addition, the Joule dissipation ($P_{J}$) in $S$ plays a role. These effects fully mask the electron cooling at low bath temperatures, where the electron-phonon thermalization is suppressed and the SINIS coolers are less efficient \cite{Giazotto2006,Pekola2004}.

\begin{figure} [t!]
    \begin{center}
    \includegraphics {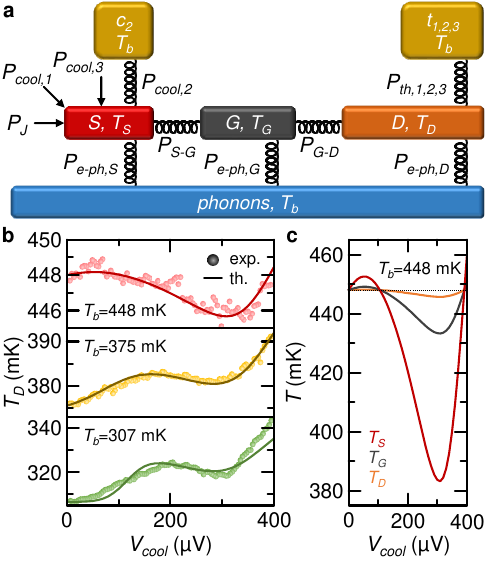}
    \end{center}
    \caption{\textbf{Thermal modeling of the device, data analysis, and graphene temperature estimation.}
    \textbf{a} Thermal model accounting for the predominant thermal exchange processes in our structure. The heat interactions between the different elements are indicated by the springs. The sign of the thermal currents depends on the operation regime: refrigeration/cooling or heating.
    \textbf{b} Experimental modulations of $T_D$ (circles) by $V_{cool}$ along with the theoretical model (lines) obtained at $T_b=448\,{\rm mK}$ (red), $T_b=375\,{\rm mK}$ (yellow) and $T_b=307\,{\rm mK}$ (green).
    \textbf{c} Electronic temperature of drain ($T_D$, orange), graphene ($T_G$, grey) and source ($T_S$, red) extracted by fitting the experimental data at $T_b=448\,{\rm mK}$. The maximum difference between the experimental value and the model of $T_D$ is 0.4 mK, which corresponds to an error of $\pm1.6$ mK on $T_G$ and of $\pm4.8$ mK on $T_S$.}
\label{Fig3}
\end{figure}

\subsection{System modelling and evaluation of the graphene electronic temperature}
To account for the experimental $T_D$ versus $V_{cool}$ characteristics and to precisely evaluate the graphene electron temperature $T_G$, we developed a thermal model accounting for all the main exchange mechanisms present in the device, as depicted in Fig.~\ref{Fig3}a. 
We assume the phonons in the different elements to be fully thermalized with the substrate since the Kapitza resistance is negligibly small at sub-Kelvin temperatures \cite{Wellstood1994}.
Furthermore, when power is extracted from (injected to) $S$ by the local coolers, the system operates in the quasi-equilibrium regime \cite{Giazotto2006,Muhonen_2012}: the quasiparticle (electron) populations are described by the Fermi distribution, the electronic temperatures of all the electrodes can be rather different ($T_S\neq T_G \neq T_D$) and out-of-equilibrium with respect to the phonons.
Thus, the steady-state temperatures $T_S$, $T_G$, and $T_D$ of our system can be calculated through the following system of energy balance equations
   \begin{subequations}
    \begin{align}
    P_{cool_1}+P_{cool_3}+P_{J}=P_{cool_2}+P_{e-ph,S}+P_{S-G}\\
    P_{S-G}=P_{e-ph,G}+P_{G-D}\\
    P_{G-D}=P_{e-ph,D}+P_{th_1}+P_{th_2}+P_{th_3},
    \label{SystemTh}
     \end{align}
    \end{subequations}
   
where Eq.~\ref{SystemTh}a describes $S$, Eq.~\ref{SystemTh}b depicts $G$ and Eq.~\ref{SystemTh}c accounts for $D$.
The electrons in $S$ are brought out-of-equilibrium by the power extracted/injected from the biased leads $c_1$ ($P_{cool_1}$) and $c_3$ ($P_{cool_3}$), and by the $I_{cool}$-induced Joule dissipation ($P_J$). These contributions are balanced by the energy exchange with the unbiased tunnel lead ($P_{cool_2}$), the phonons in $S$ ($P_{e-ph,S}$), and the electrons in the graphene flake ($P_{S-G}$). The latter is balanced in $G$ by the electron-phonon scattering ($P_{e-ph,G}$) and the electron diffusion towards $D$ ($P_{G-D}$). Finally, the drain electrons thermalize with the phonons ($P_{e-ph,D}$) and the three thermometers ($P_{th_1}$, $P_{th_2}$ and $P_{th_3}$).
Concluding, we can exploit Eqs.~\ref{SystemTh} to fit the experimental data ($T_D$) and to extract the steady-state values of $T_G$.
The details regarding the theoretical model and each contribution are presented in the Experimental Section.

Figure~\ref{Fig3}b shows the comparison between the experimental $T_D(V_{cool})$ characteristics (circles) and the theoretical curves (lines) at selected values of $T_b$. Despite its apparent complexity, our model provides an excellent agreement with the experimental data both in \emph{cooling} (see the red trace for $T_b=448\,{\rm mK}$) and \emph{refrigeration} (see the yellow and green curves for $T_b=375\,{\rm mK}$ and $T_b=307\,{\rm mK}$, respectively). Notably, our theoretical calculations exploit only data arising from the initial calibration of the tunnel junctions characteristics versus temperature (see Fig. \ref{Fig1}b and \ref{Fig1}c), without the need of further inputs. Thus, the model can be exploited to adequately infer the values of $T_G$ and $T_D$. Indeed, Fig.~\ref{Fig3}c shows the $V_{cool}$-dependence of $T_S$ (red), $T_G$ (grey) and $T_D$ (orange) obtained by substituting the experimental data acquired at $T_b=448\,{\rm mK}$ in the different contributions of Eq. \ref{SystemTh}. The temperature modulation strongly reduces from $S$ to $D$. Indeed, the electrons in $S$ are cooled of $\sim65\,{\rm mK}$, in $G$ of $\sim15\,{\rm mK}$ and in $D$ of $\sim3\,{\rm mK}$, thus only about the $4.6\%$ of the energy extracted from (injected in) $S$ is extracted from (injected in) $D$. By contrast, our device is able to modulate $T_G$ with an efficiency as high as $\sim23\%$, thereby confirming the aptness of the design of our structure as graphene electronic cooler.

\begin{figure*} [t!]
    \begin{center}
    \includegraphics {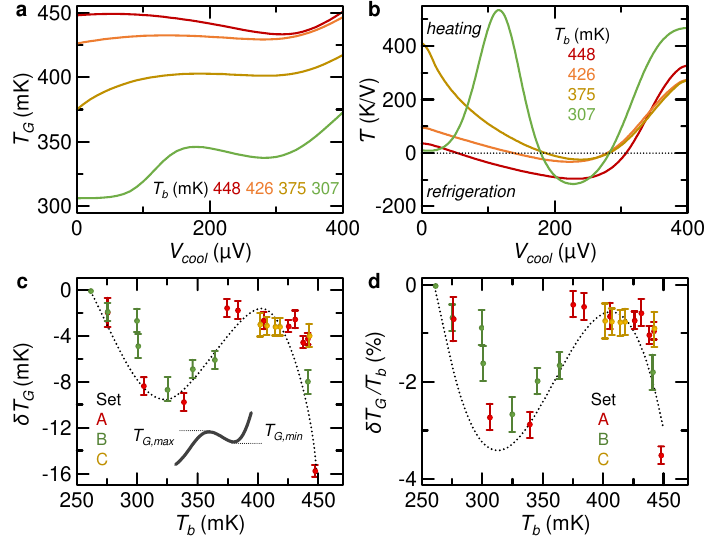}
    \end{center}
    \caption{\textbf{Electron refrigeration of graphene.}
    \textbf{a} Electronic temperature of graphene ($T_G$) versus the cooler voltage bias ($V_{cool}$) extracted for selected values of phonon temperature ($T_b$). The maximum error in the estimate of $T_G$ is $\pm2$ mK.
    \textbf{b} Voltage-to-temperature transfer function ($\mathcal{T}=\text{d}T_G/\text{d}V_{cool}$) versus cooler voltage bias ($V_{cool}$) extracted for selected values of phonon temperature ($T_b$). The maximum error in the estimate of $\mathcal{T}$ is $\pm16\,{\rm K/ V}$.
    \textbf{c} Electronic refrigeration of graphene ($\delta T_G=T_{G,min}-T_{G,max}$) versus phonon temperature ($T_b$) extracted for different data sets. The inset schematically depicts the definition of $T_{G,min}$ and $T_{G,max}$. The dotted line is a guide to the eye.
    \textbf{d} Normalized graphene electronic refrigeration ($\delta T_G/T_b$) versus $T_b$ extracted for different data sets. The dotted line is a guide to the eye.}
\label{Fig4}
\end{figure*}

\subsection{Refrigeration performance}
We now focus on the refrigeration performance of the graphene electrons as a function of voltage bias and bath temperature. 
Figure~\ref{Fig4}a shows the $V_{cool}$-dependence of $T_G$ for selected values of $T_b$. In full agreement with the measured values of $T_D$, $T_G$ increases with respect to $T_b$ at low voltage biases, while it decreases when $|V_{cool}|\lesssim(\Delta_{cool_1}+\Delta_{cool_3})/e$. By further raising the voltage bias, the tunnel electrodes act as local heaters and $T_G$ rises monotonically. This phenomenology can be effectively captured by the voltage-to-temperature transfer function, defined as $\mathcal{T}=\text{d}T_G/\text{d}V_{cool}$, which is a relevant figure of merit for active electron coolers \cite{Paolucci_2017}. Indeed, $\mathcal{T}>0$ indicates that the graphene electron temperature rises, while $\mathcal{T}<0$ defines cooling (or at least refrigeration). The device efficiency rises with $T_b$, but the range of $V_{cool}$ shrinks. In particular, the best cooling efficiency $\mathcal{T}=(-115\pm 4)\,{\rm K/ V}$ are shown at $V_{cool}=230\;\mu$V and $T_b=307\,{\rm mK}$, while the widest range of refrigeration $\delta V_{cool}\simeq300\;\mu$K occurs at $T_b=448\,{\rm mK}$.

We now analyse the graphene refrigeration in absolute terms by defining $\delta T_G=T_{G,max}-T_{G,min}$, where $T_{G,max}$ is the maximum value of the graphene temperature for $|V_{cool}|  <(\Delta_{cool_1}+\Delta_{cool_3})/e$ and $T_{G,min}$ is the minimum graphene electronic temperature for $V_{cool}\neq0$.
Figure~\ref{Fig4}c shows the dependence of $\delta T_G$ on $T_b$ for three different sets of data ($A$, $B$ and $C$). The different data show comparable results, thus confirming both the reproducibility of the thermometry experiments and the accuracy of our model. 
The raw data and theoretical modeling of data set $B$ can be found in the Supporting Information.
Beyond showing larger values of $\delta T_G$, the data sets acquired at higher bath temperatures exhibit net \emph{cooling} which reduces from $\sim 15\,{\rm mK}$ to $\sim 1\,{\rm mK}$ by decreasing $T_b$. 
Moreover, the $\delta T_G(T_b)$ characteristics show a strongly non-monotonic oscillatory behavior, since the different thermal conduction channels have completely different temperature dependencies (see the Experimental Section for details). The highest refrigeration was measured at $T_b=448\,{\rm mK}$, since the maximum efficiency of a NIS cooler occurs at $\sim0.4T_C$ \cite{Giazotto2006} ($T_b=448\,{\rm mK}$$\simeq0.39T_{C,cool_1}\simeq0.36T_{C,cool_2}$).
Finally, we compare $\delta T_G$ with the initial electronic temperature by defining the normalized graphene electronic refrigeration $\delta T_G/T_b$ shown in Fig.~\ref{Fig4}d. The maximum normalized refrigeration is about 3.5$\%$ at $T_b=448$ mK and it re-increases to $\sim 3\%$ in the range $300$ mK$\leq T_b\leq330$ mK, pointing out that our graphene electron refrigeration can be achieved in a wide range of temperatures.

\section{Conclusions}
In conclusion, we demonstrated the possibility of actively cooling the graphene electrons under the crystal temperature in cryogenic environments. To this scope, we designed and realized a structure exploiting superconducting tunnel junctions operating as both coolers and thermometers. 
Thanks to the \emph{non-local} cooling geometry and the good thermal diffusion over the whole device, we were able to avoid the strong Joule dissipation produced by electronic currents flowing in graphene and to circumvent the limitations arising from the low quality of the superconducting tunnel junctions directly realized on graphene \cite{Bretheau2017}.
In our experiment, the graphene temperature was calculated thanks to a dedicated model accounting for all the main thermal exchange mechanisms present in the system showing a very good agreement with the measured values of the drain temperature. Interestingly, we demonstrated refrigeration of the graphene electrons by simple voltage in a wide range of bath temperatures ($250-450\,{\rm mK}$). 
Furthermore, our device also showed a net \emph{cooling} of the graphene electrons from the starting phonon temperature with a maximum of $(15.5\pm0.5)\,{\rm mK}$ at $T_b=448\,{\rm mK}$ and a best voltage-to-temperature transfer function of $(-115\pm4)\,{\rm K/ V}$ at $T_b=307\,{\rm mK}$.
Within the same geometry, the graphene refrigeration efficiency could be boosted by decreasing the normal-state resistance of the local tunnel coolers and by decreasing the size of the $\Gamma$ broadening parameter (improved fabrication quality). Further improvements could be obtained by equipping the superconducting cooler with quasiparticle traps \cite{Nguyen_2013} or by suspending the graphene channel to bring out-of-equilibrium also its phonons \cite{Koppinin2009}. Moreover, since $D$ was only needed to demonstrate the heat diffusion over the whole structure, the drain electrode could be removed in practical applications to decrease the total thermal losses and increase the cooling performance.

Despite directly lowering and measuring the graphene electronic temperature by implementing superconducting tunnel electrodes coupled to $G$ are theoretically possible \cite{Vischi2020}, such a configuration comes with some experimental drawbacks. First, practical superconductor/graphene tunnel junctions exploiting 2D crystals as dielectric material, such as hexagonal boron nitride (hBN)\cite{Bretheau2017} or molybdenum disulfide \cite{Dvir2018-ox}, show a lower quality of charge transport (high value of $\Gamma$) compared to fully metallic systems. Indeed, the $\Gamma$ decreases the cooling efficiency of the system \cite{Giazotto2006}, thus our non-local configuration shows superior performance. Second, the tunnel resistivity of insulating 2D crystals is rather large and grows exponentially with the number of layers \cite{Yu2011,Roy2015-ky,DUTTA20241}. Thus, the cooling efficiency of the superconducting tunnel probe decreases, since the cooling power is proportional to $1/R$ \cite{Giazotto2006}. Differently, tunnel junctions fabricated by oxidizing the metallic thin film can reach values $\sim$1 k$\Omega$ \cite{Nahum1994}, thus the performance of our device are superior. Third, the charge current necessary to operate tunnel coolers and thermometers would create a sizeable Joule overheating in $G$, due to the large graphene resistance (normal metal islands have usually resistances of a few ohms), thus compromising the overall cooling efficiency. Consequently, despite the complex structure, our \emph{non-local} configuration promises larger modulations of $T_G$ compared to direct graphene cooling approaches.

Our structure could enable the design and realization of innovative devices in the realm of hybrid graphene /superconductor quantum technology. 
Indeed, our work is the first step towards the realization of graphene cold electron bolometers \cite{Vischi2020}, which might also benefit from our non-local configuration (both in cooling and read-out) to avoid the strong Joule dissipation due to the high resistivity of graphene and the poor cooling efficiency of the superconductor/graphene tunnel junctions.
Furthermore, the graphene electron coolers equipped with superconducting electrodes made of lower temperature superconductors, such as molybdenum ($T_C\sim920$ mK) and titanium ($T_C\sim400$ mK), could be exploited to increase the lifetime of transmon qubits \cite{Wang2019-uf,Kroll2018-ue} or improve the efficiency of Josephson parametric amplifiers \cite{Butseraen2022-gv}.
Finally, graphene might be a valid material to increase the tunability and versatility of phase-coherent caloritronics \cite{Fornieri2017-bn}, since it integrates the control thorough the superconducting phase with the possibility to master the thermal properties of graphene by gating.


\section{Experimental Section}
\subsection{Nano-fabrication}
The device was fabricated onto a $8\times8\,{\rm mm^2}$ chip obtained by dicing an intrinsic silicon wafer covered with $300\,{\rm nm}$ of thermally grown dry silicon dioxide (SiO$_2$).
To locate the flakes onto the chip, metallic markers were fabricated through electron-beam lithography (EBL, Zeiss UltraPlus with Raith Elphy Multibeam patterning system) followed by thermal evaporation of $5\,{\rm nm}$ of Ti and $35\,{\rm nm}$ of Au.
Then, the chip was cleaned by oxygen plasma at $100\,{\rm W}$ for 5 minutes to remove organic residues on its surface.\\
Graphene flakes were deposited on the substrate by micro-mechanical exfoliation of natural graphite (NGS Trade and Consulting GmbH) by means of blue tape (Nitto Italia srl). Indeed, this technique is a simple and cheap method to obtain almost defect-free graphene flakes of lateral dimensions suitable for our devices \cite{Liu2024-yu}.
Monolayer flakes were identified by optical contrast method \cite{contrast,RodGra} by using a motorized stage optical microscope (DM12000 M, Leica). Then, Raman spectroscopy (InVia spectrometer, Renishaw) was used to confirm the graphene thickness and the absence of defects (see Supporting Information).
The $A_G=8\times3\,{\rm \mu m^2}$ graphene channel of the device was shaped by EBL and reactive ion etching (RIE).\\
The electrodes were fabricated by a single EBL step and two-angle shadow-mask metals deposition through a suspended bilayer resist mask ($\sim700\,{\rm nm}$-thick MMA film covered by a $250\,{\rm nm}$-thick PMMA layer) \cite{Dolan}. This technique relies on the strong homogeneity of the resist thin film thickness, thus the spin coating procedure is conventionally performed on an entire round wafer. Indeed, a square substrate shows an increase of the spin coated resist at the rim (especially at the corners) comparable with its total thickness \cite{coatings11111322}.
In our fabrication process, the suspended mask was realized on a small substrate by processing graphene flakes in the center of the chip. To have homogeneous films, we employed a rotation speed of 4000 rpm for both layers (the conventional rotation speed for spin-coating of MMA on 2 inch wafers is $\sim800$ rpm). 
As a result, the central $5\times5$ mm$^2$ of the chip are almost homogeneous (a
variation $\pm20$ nm), since the higher thickness at the rim propagates for several hundred of micrometers \cite{coatings11111322}. These total roughness of the MMA/PMMA bilayer generates an error of the position of the superconducting cooler of $\pm2.5$ nm, which does not affect the operation performance of the device (see Supporting Information for further details). Conversely, the MMA film is thinner than usual ($\sim$1.2 $\mu$m), thus allowing for lower shift of the different components of the structure.
The evaporation and oxidation processes were performed in an ultra-high vacuum electron-beam evaporator of base pressure $10^{-11}\,{\rm Torr}$ (DCA). At first, $3\,{\rm nm}$ of Ti and $15\,{\rm nm}$ of Al$\textsubscript{0.98}$Mn$\textsubscript{0.02}$ were deposited at 0$^\circ$ to realize the $S$ and $D$ electrodes (deposition rate $\sim1.5\,$\AA/s). 
The volumes of the Ti and Al$\textsubscript{0.98}$Mn$\textsubscript{0.02}$ component of $S$ ($D$) are $\mathcal{V}_{S,Ti}\simeq2.1\times10^{-3}\,{\rm\mu m^3}$ ($\mathcal{V}_{D,Ti}\simeq1.3\times10^{-3}\,{\rm\mu m^3}$) and $\mathcal{V}_{S,AlMn}\simeq1.05\times10^{-2}\,{\rm\mu m^3}$ ($\mathcal{V}_{D,AlMn}\simeq6.6\times10^{-3}\,{\rm\mu m^3}$), respectively. 
Subsequently, the sample was exposed to $200\,{\rm mTorr}$ of $\textrm O_{2}$ for 5 min in order to realize the AlO$_x$ thin layer forming the tunnel barriers. Finally, a $35\,{\rm nm}$-thick Al layer was deposited at a tilt angle of $40^\circ$ (deposition rate $\sim2\,$\AA/s) to form the superconducting coolers ($c_1$, $c_2$ and $c_3$) and thermometers ($t_1$, $t_2$ and $t_3$).
The large angle difference between the two metal depositions allowed us to compensate for the small inhomogeneity of the MMA layer, thus avoiding unwanted superpositions of different parts of the device.

\subsection{Measurements}
All experiments were performed in a filtered He$^3$-He$^4$ dry dilution refrigerator (Triton 200, Oxford Instruments) at bath temperatures ranging from $\sim90\,{\rm mK}$ to $\sim500\,{\rm mK}$.\\
The charge transport properties of all the tunnel junctions were recorded at different phonon bath temperatures in a standard two-wire configuration by applying a voltage bias through a floating source (GS200, Yokogawa), and by measuring the current with a room-temperature current pre-amplifier (Model 1211, DL Instruments). By employing the same set-up, we measured the charge transport through graphene to ensure the good electron diffusion across the entire device. \\ 
The thermal transport experiments were performed by current biasing the SINIS thermometer\cite{Rowell1976} embodied in $D$ by means of a custom-made low-noise floating source ($I_{th}=300\,{\rm pA}$) and monitoring the voltage drop ($V_{th}$) with a standard room temperature voltage pre-amplifier (Model 1201, DL Instruments). The electronic temperature of $S$ was tuned by voltage biasing a couple of tunnel coolers by a floating source (GS200, Yokogawa). The calibration of the SINIS thermometer was realized by measuring $V_{th}$ at $I_{th}=300\,{\rm pA}$ while slowly varying the cryostat temperature. 

\subsection{Device parameters}
In this section we provide the basics parameters of our device. All the reported quantities were measured by Raman spectroscopy, charge transport experiments and thermal experiments. \\
The normal-state resistance of the three cooler tunnel junctions is $R_{cool_1}\simeq3.6\,{\rm k}$$\Omega$, $R_{cool_2}\simeq22\,{\rm k}$$\Omega$ and $R_{cool_3}\simeq17.8\,{\rm k}$$\Omega$, respectively. The values of the normal-state resistance of the three coolers arises from the different areas of the tunnel junctions (in particular $c_1$ has the larger area).
The zero-temperature energy gap of the superconducting coolers is $\Delta_{0,cool_1}\simeq175\,{\rm \mu eV }$, $\Delta_{0,cool_2}\simeq190\,{\rm \mu eV }$ and $\Delta_{0,cool_3}\simeq190\,{\rm \mu eV }$. Consequently, the critical temperature of the three leads is
$T_{C,cool_1}\simeq1.15\,{\rm K}$, $T_{C,cool_2}\simeq1.25\,{\rm K}$ and $T_{C,cool_3}\simeq1.25\,{\rm K}$.
Finally, the Dynes parameter accounting for sub-gap quasiparticle population \cite{Dynes1984} is $\Gamma_{cool_1}=\Gamma_{cool_2}=\Gamma_{cool_3}=2\times 10^{-2}$.\\
The normal-state resistance of the thermometer tunnel junctions is $R_{th_1}\simeq26.1\,{\rm k}$$\Omega$, $R_{th_2}\simeq25\,{\rm k}$$\Omega$ and $R_{th_3}\simeq26\,{\rm k}$$\Omega$, respectively. The zero-temperatures energy gap of the superconducting thermometers is $\Delta_{0,th_1}\simeq200\,{\rm \mu eV }$, $\Delta_{0,th_2}\simeq200\,{\rm \mu eV }$ and $\Delta_{0,th_3}\simeq200\,{\rm \mu eV }$. 
The resulting critical temperature of the three probes is
$T_{C,th_1}=T_{C,th_2}=T_{C,th_3}\simeq1.32\,{\rm K}$.
Finally, the Dynes parameter is $\Gamma_{th_1}=\Gamma_{th_2}=\Gamma_{th_3}=2\times 10^{-2}$.\\
By measuring devices fabricated with the same procedure, we estimated the graphene mobility to be $\mu_G\simeq10^3$ cm$^2$/(V$\cdot$s). The graphene charge carrier concentration $n_G\simeq3\times10^{12}\,{\rm cm^{-2}}$ was estimated from the Raman spectra (see Supporting Information). Consequently, the contact resistances of the graphene with the $S$ and $D$ are $R_{S-G}\simeq15\,{\rm k\Omega}$ and $R_{G-D}\simeq15\,{\rm k\Omega}$, respectively.\\
We note that the measurement error of the all the device parameters is lower than 1$\%$. Thus, the measured value of $T_D$ is affected by the same error, which provides a negligible contribution to the estimate of $T_G$ and $T_S$.

\subsection{Theoretical Model}
The $V_{cool}$-dependent experimental values of $T_D$ were extracted for different $T_b$ by exploiting the theoretical model accounting for the main thermal exchange mechanisms present in the device shown in Eq.~\ref{SystemTh}. Here we describe in detail all the contributions.\\
The energy transfer between a metallic island and a tunnel-coupled superconductor can be written \cite{Giazotto2006}
\begin{equation}
P_k(T_i,T_j)=\frac{1}{e^2R_k}\!\int_{-\infty}^{+\infty}\!\!\!\!\!\!\!\!\!\!d\epsilon \, \epsilon N_{i}(\epsilon )[f(\epsilon,T_i)-f(\epsilon+eV,T_j)],
\label{eq:Ptunn}
\end{equation}
where $e$ is the electron charge, $R_k$ is the normal-state resistance of the $k$ tunnel junction, $V$ is the voltage drop across the junction, $N_i(\epsilon)$ is the quasiparticle density of states (DoS) of the superconductor, while $f(\epsilon,T_i)$ and $f(\epsilon+eV,T_j)$ functions of the superconducting $i$ and normal metal $j$ electrode, respectively. 
The superconducting $i$-lead shows a smeared BCS DoS $N_i(\epsilon)=|\text{Re}[(\epsilon+i\Delta_{0,i}\Gamma_i)/\sqrt{(\epsilon+i\Delta_{0,i}\Gamma_i)^2-\Delta_i(T_i)^2}]|$ \cite{tinkham2004}, where $\Delta_i(T_i)$ is the temperature-dependent energy gap, $\Delta_{0,i}$ the zero-temperature energy gap and $\Gamma_i$ is the Dynes parameter \cite{Dynes1984}.\\
The power extracted/injected by $cool_1$ ($cool_3$) from/to $S$ $P_{cool_1}$ ($P_{cool_3}$) can be computed by substituting $R_{cool_1}$ ($R_{cool_3}$), $\Delta_{0,cool_1}$ ($\Delta_{0,cool_3}$), $\Gamma_{cool_1}$ ($\Gamma_{cool_3}$) and $V=V_{cool}/2$ ($V=V_{cool}/2$) in Eq.~\ref{eq:Ptunn}. Differently, $P_{cool_2}$ is calculated by exploiting $V=0$, $R_{cool_2}$, $\Delta_{0,cool_2}$ and $\Gamma_{cool_2}$, since $cool_2$ was not biased.\\
The drain electrode is tunnel coupled to two biased and one unbiased superconducting thermometers. Thus, beyond the constructive parameters of the three junctions, we evaluate $P_{th_1}$ and $P_{th_3}$ at $V=V_{th}/2$, while $P_{th_2}$ is computed at $V=0$.\\
The current flowing in $S$ due to the voltage bias applied to the coolers produced a Joule power dissipation 
\begin{equation}
    P_{Joule}(T_S)=R_SI_{cool}^2(T_S),
\end{equation}
where $I_{cool}(T_S)$ is the $T_S$-dependent experimental value of the charge current flowing through the two biased coolers (see Fig.~\ref{Fig1}b) and $R_S$ is the resistance of the portion of $S$ between the tunnel coolers. The latter is estimated as $R_S=\frac{l_S}{w_St_S}\frac{1}{e^2D_{AlMn}N_{AlMn}}=53.7\;\Omega$, with $l_S=1.5\;\mu$m the distance between the two coolers, $w_S=150$ nm the width of the two coolers, $t_S=15$ nm the film thickness, $D_{AlMn}=2.25\times10^{-3}$ m$^2$s$^{-1}$ the AlMn diffusion coefficient and $N_{AlMn}=2.15\times10^{47}$ m$^{-3}$J$^{-1}$ the density of states at the Fermi energy of AlMn \cite{Fornieri2014-de}. We consider only the conductivity of the AlMn film since the Ti layer acts only as an adhesive layer and does not contribute to the charge transport.\\
The electron-phonon thermal exchange in the $S$ and $D$ electrodes is described by \cite{Giazotto2006}
\begin{equation}
\begin{split}
    P_{e-ph,i}(T_i,T_b)=\\
    \mathcal{V}_{i,Ti}\Sigma_{Ti}\big( T_i^5-T_b^5\big)+\mathcal{V}_{i,AlMn}\Sigma_{AlMn}\big( T_i^6-T_b^6\big),
\end{split}  
\end{equation}
where $\Sigma_{Ti}=1.3\times10^9$ W$\,$m$^{-3}$K$^{-5}$ is the electron-phonon coupling constant of Ti, $\Sigma_{AlMn}=4.5\times10^9$ W$\,$m$^{-3}$K$^{-6}$ is the electron-phonon scattering constant of Al$\textsubscript{0.98}$Mn$\textsubscript{0.02}$ \cite{Giazotto2006} and $i=S,D$. Indeed, the total electron-phonon heat exchange in a metallic thin bilayer is exactly given by the sum of the contributions of the separate components \cite{Paolucci2020}.\\
The thermal exchange between graphene ($G$) and the metal electrodes ($S$ and $D$) can be described by \cite{Giazotto2006}
\begin{equation}
    P_{i-j}(T_i,T_j)=\frac{\pi^2}{6R_{i-j}}\bigg(\frac{k_B}{e}\bigg)^2 \big(T_i^2-T_j^2\big),
\end{equation}
where $i=S$ and $j=G$ for $P_{S-G}$, while $i=G$ and $j=D$ for $P_{G-D}$.\\
The electron-phonon thermalization in diffusive graphene is given by \cite{Chen2012}
\begin{equation}
    P_{e-ph,G}(T_G,T_b)=A_G\Sigma_G\big( T_G^3-T_b^3\big).
\end{equation}
Here, the electron phonon coupling constant can be written $\Sigma_G=\frac{2\zeta(3)D_P^2k_B^3e}{\pi^2\rho_M\hbar^4v_F^2v_s^2\mu_G}$, where $\zeta(3)\simeq1.2$ is the Apery constant, $D_P=13\,{\rm eV}$ is the deformation potential, $\rho_M=7.6\times 10^{-7}\,{\rm kg/m^2}$ is the mass density, $v_F=10^6\,{\rm m/s}$ is the Fermi velocity and $v_s=2\times 10^4\,{\rm m/s}$ is the sound velocity of graphene \cite{CastroNeto2009}. Notably, $\Sigma_G$ is independent of the charge carrier concentration ($n_G$), thus decreasing the uncertainty on its value.\\
To extract $T_G$, we exploit only measured quantities (from charge or thermal transport measurements) without using any additional fitting parameter. Thus, the accuracy of the fit of the experimental $T_D$ versus $V_{cool}$ characteristics ensures the validity of the thermal model of the system, as shown in Fig.~\ref{Fig3}b for different values of $T_b$.

\medskip
\textbf{Acknowledgements} \par 
The authors thank E. Mangiacotti for fruitful discussions and Dr. C. Coletti from the Istituto Italiano di Tecnologia for the access to the micro-Raman facility.
F. B. and F. P. acknowledge partial financial support under the National Recovery and Resilience Plan (NRRP), Mission 4, Component 2, Investment 1.1, Call PRIN 2022 by the Italian Ministry of University and Research (MUR), funded by the European Union – NextGenerationEU – EQUATE Project, "Defect engineered graphene for electro-thermal quantum technology" - Grant Assignment Decree No. 2022Z7RHRS.
F. G. acknowledges the EU's Horizon 2020 Research and Innovation Framework Programme under the grant No. 964398 (SUPERGATE) and No. 101057977 (SPECTRUM) for partial financial support.  S.R. acknowledges the financial support of the TRUST project (PRIN 2022M5RSK5). 

\textbf{Conflict of interest.}
The authors declare no conflict of interest.

\medskip

%
\providecommand{\latin}[1]{#1}
\makeatletter
\providecommand{\doi}
  {\begingroup\let\do\@makeother\dospecials
  \catcode`\{=1 \catcode`\}=2 \doi@aux}
\providecommand{\doi@aux}[1]{\endgroup\texttt{#1}}
\makeatother
\providecommand*\mcitethebibliography{\thebibliography}
\csname @ifundefined\endcsname{endmcitethebibliography}  {\let\endmcitethebibliography\endthebibliography}{}


\begin{figure}
\textbf{Table of Contents}\\
\medskip
  \includegraphics{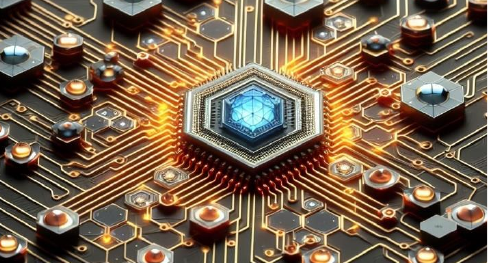}
  \medskip
  \caption*{ToC Entry}
\end{figure}

\newpage
\newpage
\newpage
\newpage
\newpage
\newpage

\section*{Supporting Information: Active electron cooling of graphene}




\section{Raman spectroscopy of graphene}

\begin{figure}[h!]
    \begin{center}
    \includegraphics {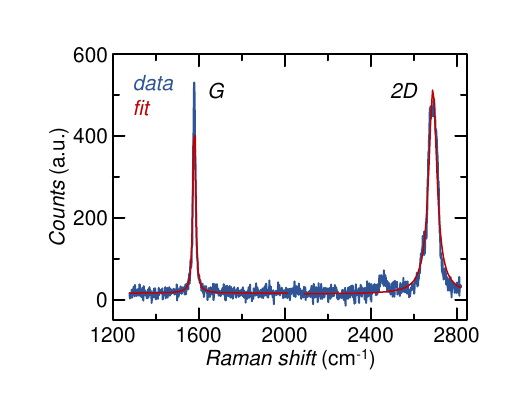}
    \end{center}
    \caption{\textbf{Raman spectroscopy.}
    Typical Raman spectrum of the graphene channel of our thermal transistor (blue) along with the Lorentian fit (red) of the $G$ and $2D$ peaks. The position of the two Raman modes is $\sim1582.7$ cm$^{-1}$ and $\sim2676.5$ cm$^{-1}$ for the $G$ and $2D$ peak, respectively. By constructing the related correlation plot \cite{Lee2012-rg}, we extracted a small strain ($\sim0.1\%$) and a charge concentration $n_G\simeq3\times10^{12}\,{\rm cm^{-2}}$ for the graphene channel.}
\label{FigS1}
\end{figure}

\newpage
\newpage
\section{Fabrication of superconducting tunnel electrodes on graphene flakes}

\begin{figure} [h!]
    \begin{center}
    \includegraphics {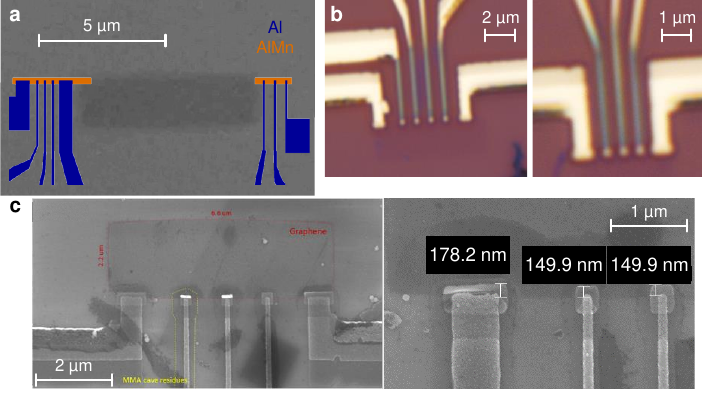}
    \end{center}
    \caption{\textbf{Fabrication steps and precision of hybrid graphene/superconductor devices with angle resolved shadow mask technique.}
    \textbf{a} Optical picture of the graphene flake used in the channel of our device with superimposed the CAD of the source and drain electrodes (orange, AlMn) and the superconducting tunnel probes (blue, Al).
    \textbf{b} Optical pictures of the test devices where the graphene flakes are contacted by small AlMn islands equipped with Al tunnel probes in a single steps of lithography by exploiting the angle resolved shadow mask technique \cite{Dolan}.
    \textbf{c} SEM pictures of the same same devices in b. The maximum misalignment between the AlMn islands and the graphene is $25$ nm, since the overlap from design is 175 nm. In addition, the experimental variability of the displacement of the Al tunnel probes of the different electrodes evaporated at an angle of 40$^\circ$ is $5$ nm. The theoretical error in the displacement of lithographed structures for the thickness of the resist (h=950 nm), the experimental roughness of 20 nm and evaporation angle employed in these devices is $\pm6.5$ nm on the ideal value of 611 nm. The experimental error is smaller than the theroretical one, because the variation of the resist thickness occurs on a lateral length scale much larger than the distance between the electrodes on the same device. Finally, these errors are compatible with the fabrication of hybrid graphene/superconductor devices.}
\label{FigS3}
\end{figure}
\newpage
\section{Data set B}

\begin{figure} [h!]
    \begin{center}
    \includegraphics {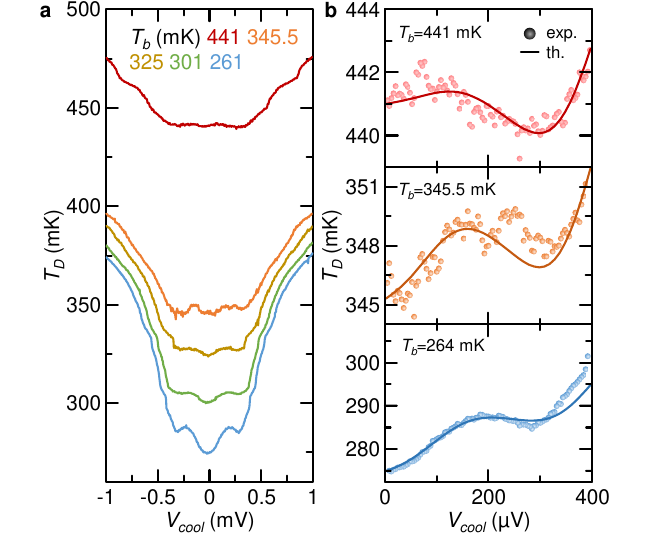}
    \end{center}
    \caption{\textbf{Temperature modulation of the drain electrode for data set $B$.}
    \textbf{a} Electronic temperature of the drain electrode ($T_D$) as a function of the voltage bias of the source coolers ($V_{cool}$) recorded at selected values of phonon temperature ($T_b$).
     \textbf{b} Experimental modulations of $T_D$ (circles) by $V_{cool}$ along with the theoretical model (lines) obtained at $T_b=441\,{\rm mK}$ (red), $T_b=345.5\,{\rm mK}$ (orange) and $T_b=264\,{\rm mK}$ (turquoise).}
\label{FigS2}
\end{figure}


\end{document}